# Observation of Unusual Topological Surface States in Half-Heusler Compounds *Ln*PtBi (*Ln*=Lu, Y)

Z. K. Liu[1,2], L. X. Yang[3], S.–C. Wu[4], C. Shekhar[4], J. Jiang[1,5], H. F. Yang[6], Y. Zhang[5], S.–K. Mo[5], Z. Hussain[5], B. Yan[1,2,4], C. Felser[4] and Y. L. Chen[1,2,3,7]

[1]School of Physical Science and Technology, ShanghaiTech University, Shanghai 200031, P. R. China
[2]CAS-Shanghai Science Research Center, 239 Zhang Heng Road, Shanghai 201203, P. R. China
[3]State Key Laboratory of Low Dimensional Quantum Physics, Department of Physics and Collaborative Innovation Center for Quantum Matter, Tsinghua University, Beijing 100084, P. R. China
[4]Max Planck Institute for Chemical Physics of Solids, D-01187 Dresden, Germany
[5]Advanced Light Source, Lawrence Berkeley National Laboratory, Berkeley, CA 94720, USA
[6]State Key Laboratory of Functional Materials for Informatics, SIMIT, Chinese Academy of Sciences, Shanghai 200050, P. R. China
[7]Physics Department, Oxford University, Oxford, OX1 3PU, UK

**Topological quantum materials represent a new class of matter with both exotic physical phenomena and novel application potentials[1,2]. Many Heusler compounds, which exhibit rich emergent properties such as unusual magnetism, superconductivity and heavy fermion behaviour[3,4,5,6,7,8,9], have been predicted to host non-trivial topological electronic structures[10,11,12,13]. The coexistence of topological order and other unusual properties makes Heusler materials ideal platform to search for new topological quantum phases (such as quantum anomalous Hall insulator and topological superconductor). By carrying out angle-resolved photoemission spectroscopy (ARPES) and *ab initio* calculations on rare-earth half-Heusler compounds *Ln*PtBi (*Ln*=Lu, Y), we directly observed the unusual topological surface states on these materials, establishing them as first members with non-trivial topological electronic structure in this class of materials. Moreover, as *Ln*PtBi compounds are non-centrosymmetric superconductors, our discovery further highlights them as promising candidates of topological superconductors.**

Topological quantum materials, a new class of matter with non-trivial topological electronic structures, has become one of the most intensively studied fields in physics and material science due to their rich scientific significance and broad application potentials[1, 2]. With the world wide effort, there have been numerous materials predicted and experimentally confirmed as topologically non-trivial matter (including topological insulators[14, 15], topological crystalline insulators[16, 17] and 3D topological Dirac and Weyl semimetals[18, 19, 20, 21, 22]). However, there is a big family of materials - the Heusler compounds - although being theoretically predicted to be topologically non-trivial back in 2010[10, 11, 13], the non-trivial topological nature has never been experimentally confirmed up to date.

The ternary semiconducting Heusler compounds, with their great diversity (~500 members, >200 are semiconductors) give us the opportunity to search for optimized parameters (e.g., spin-orbit coupling (SOC) strength, gap size, etc.) across different compounds - which is critical not only for realizing the topological order and investigating the topological phase transitions[10], but also for designing realistic applications. In addition, among the wealth of Heusler compounds, many (especially those containing rare-earth elements with strongly correlated $f$-electrons) exhibit rich interesting ground state properties, such as magnetism[4, 23], superconductivity[5, 6, 8] or heavy fermion behaviour[3]. The interplay between these properties and the topological order makes Heusler compounds ideal platforms for the realization of novel topological effects (e.g. exotic particles including image monopole effect and axions, etc.), new topological phases (e.g. topological superconductors[24, 25]) and broad applications (see ref. [26] for a review).

Rare-earth half-Heusler compounds $Ln$PtBi ($Ln$=Y, La and Lu) represent a model system recently proposed that can possess topological orders with nontrivial topological surface states (TSSs) and large band inversion[10, 27]. Moreover, due to the lack of the inversion symmetry in their crystal structure, non-centrosymmetric superconducting $Ln$PtBi compounds (Tc=0.7[5], 0.9[6] and 1.0 K[8] for $Ln$=Y, La and Lu, respectively) may also host unconventional cooper pairs with mixed-parity, making them promising candidates for the investigation of topological superconductivity and the search for Majorana fermions[28].

However, despite the great interests and intensive research efforts in both theoretical[10, 11, 12] and experimental[9, 29, 30] investigations, the topological nature on *Ln*PtBi remains elusive. A previous ARPES study has reported metallic surface states[29] with apparently different dispersion shape and Fermi surface crossing numbers from the predicted TSSs in *Ln*PtBi compounds[10, 11], making the topological nature of *Ln*PtBi controversial.

In this work, by carefully performing comprehensive ARPES measurements and *ab-initio* calculations, we resolved this unsettled question. For the first time, we observed the non-trivial TSSs with linear dispersions in half-heusler compounds LuPtBi and YPtBi (on the Bi- and Y-terminated (111) surface, respectively); and remarkably, in contrast to many topological insulators (TIs) that have TSSs inside their bulk gap[1, 14, 31], the TSSs in *Ln*PtBi show their unusual robustness by lying well below the Fermi energy ($E_F$) and strongly overlapping with the bulk valence bands (similar to those in HgTe[32, 33, 34]). In addition to the TSSs, we also observed numerous metallic surface states (SSs) crossing the $E_F$ with large Rashba splitting, which not only makes them promising compounds for spintronic application, but also provides the possibility to mediate topologically non-trivial superconductivity in the superconducting phase of these compounds.

A crystallographic unit cell of *Ln*PtBi is shown in Fig. 1a, which comprises of a zinc-blend unit cell from Bi and Pt atoms and rocksalt unit cell from Bi and *Ln*. For ARPES measurements, the *Ln*PtBi single crystals were cleaved *in-situ* in the ultra-high vacuum (UHV) measurement chamber, resulting in either (111) or (001) surfaces. The unit cell along the (111) cleavage surface is illustrated in Fig. 1b and the corresponding hexagonal surface Brillouin zone (BZ) is shown in Fig. 1d, which could be viewed as the projection of the bulk BZ (Fig. 1c) of *Ln*PtBi along the [111] direction (Fig. 1d).

The unit cell along the [111] direction consists of alternating *Ln*, Pt and Bi layers (Fig. 1b). As there are fewer chemical bonds to break between *Ln*-Bi layer (2 comparing to 3 between *Ln*-Pt or

Pt-Bi layers) and the *Ln*-Bi layer distance is twice as large as the *Ln*-Pt or Pt-Bi layer spacing (see Fig. 1b), it is more energetically favourable to cleave the material between *Ln*-Bi layers. In the discussion of the main text, we'll focus on the electronic structure of Bi-terminated (111) surface LuPtBi and Y-terminated (111) surface of YPtBi. ARPES results (as well as *ab-initio* calculations) along (001) cleavage surfaces of LuPtBi and YPtBi are presented in the Supplemental Materials.

The core level photoemission spectra of LuPtBi is shown in Fig. 1e, from which the characteristic Bi *5d* and Lu *4f* doublets are clearly observed. The large spectral weight of Bi peaks over the Lu peaks in the (111) surface clearly indicates its Bi-termination. The broad area Fermi surface (FS) mapping covering multiple Brillouin zones (BZs) in Fig. 1f also illustrates the hexagonal symmetry (with the correct lattice parameters) resulting from the (111) cleaved surface.

In Fig. 2, detailed electronic structures of LuPtBi within a surface BZ are illustrated. From the FS maps (Fig. 2a-c) and 3D spectral intensity plots (Fig. 2d,e) around both the $\bar{\Gamma}$ point and BZ boundary ($\bar{K}$ and $\bar{M}$ points), there are clearly multiple bands crossing $E_F$, forming a twin hexagonal hole pockets at $\bar{\Gamma}$ and complex electron pockets at $\bar{K}$ and $\bar{M}$, both of which show clear Rashba splitting. Around $\bar{\Gamma}$, there is another pair of double 'Λ' shape hole bands just below $E_F$. These features broadly agree with the previous ARPES report[29]. However, in this work, with the high instrument resolution and data statistics we successfully observed a critical additional 'X' shape band dispersing between 0.4 eV and 0.8 eV with the band crossing point (i.e. Dirac point) at $E_b \sim$ 0.5 eV at the $\bar{\Gamma}$ point – which is the long-sought-after topological surface states, as we will discuss in details below.

To help understand the origin of these electronic states, we carried out band structure calculations of Bi-terminated LuPtBi (111) surface with two different methods (Fig. 3a, b, see the Methods section and Supplementary Information for more details), and both agree well with the measurement and clearly reproduce the 'X' shape TSS observed in our measurements (Fig. 3c-f). In Fig. 3a, we first employed a slab model for the *ab-initio* calculations (method one). This method, which takes into account the charge density redistribution due to surface potential modification by *ab-initio* calculations, can describe all surface states including TSSs and those from the trivial dangling bonds. To further identify the TSS, we carried out another method (Fig. 3b) by calculating the k-resolved local density of states (LDOS) of a semi-infinite surface using the recursive Green's function (method two)[35] constructed from Wannier function-based tight-binding parameters extracted from the bulk material[36]. Such method could reveal TSSs clearly and avoid the trivial surface states by removing the dangling bond orbitals in the Hamiltonian (as can be clearly seen in Fig. 3b).

The combination of the two methods thus allows us to unambiguously identify the TSSs from other trivial dangling bond states. As shown in Fig. 3a, all sharp dispersions (three Kramers pairs and one 'X' shape state, labelled as SS1-SS3 and TSS respectively) are of surface origin and agree excellently with the observed band structures (Fig. 3c-f). Moreover, the three Kramers pairs (SS1, SS2, SS3) are all absent in the result from method two (Fig. 3b) while the 'X' shape band remains, illustrating their topologically trivial origin (i.e. being trivial surface states due to dangling bonds), as opposite to the TSS shown in Fig. 3b.

The surface origin of both the TSS and SS1-SS3 in Fig. 3 can also be experimentally verified by performing the photon energy dependence photoemission measurement[31], as presented in Fig. 4. In Fig. 4a, dispersions along $\bar{\Gamma}$-$\bar{K}$-$\bar{M}$ directions measured using a wide range of photon energies (50 ~ 75 eV) were plotted, the dispersions of TSS and SS1~SS3 under all photon energies are identical (though the relative intensity can vary with photon energy due to the photoemission matrix element effect[31]). To further visualize the dispersion of these bands along $k_z$, we extract the momentum distribution curves (MDCs) at $E_F$ (cutting through SS1 and SS2) and 0.65 eV below $E_F$ (cutting through TSS and SS2, SS3) and plot them as the function of photon energy (see Fig. 4b,c). Evidently, the peaks from TSS and SS1-SS3 bands show no $k_z$-dispersion as they all form straight vertical lines. Thus the surface nature of these bands (TSS and SS1~SS3) are clearly established.

By fitting the Dirac type 'X' shape linear dispersion (Fig. 4d), we can extract the band velocity at the Dirac point as 2.37 eV•Å ($3.59\times10^5$ m/s) and 3.13 eV•Å ($4.74\times10^5$ m/s) along the $\bar{\Gamma}$-$\bar{K}$ and $\bar{\Gamma}$-$\bar{M}$ direction, respectively.

Similarly, for the other compound YPtBi, our calculation and measurements also agree well and both clearly show the TSS (Fig. 4e-g, also see Supplemental Materials for the calculation). More measurements, as well as calculations on different cleaved surfaces (001) and different termination layers (Bi- or *Ln* terminations), are presented in Supplemental Materials, all showing excellent agreements. Interestingly, in both compounds, the observed TSS coexist and overlap in energy with the bulk valence band (appear as broad dispersing background intensity in Fig. 3, 4), demonstrating its unusual robustness.

Our discovery of the unusual topological surface states on these materials establishing them as first examples with non-trivial topological electronic structure showing unusual robustness in Heusler materials with great tenability due to the vast number of compounds in the family. Moreover, the interplay between topological electronic structure and the rich properties in Heusler materials further makes them ideal platforms for the realization of novel topological effects (e.g. exotic particles including image monopoles[37] and axions[38]) and new topological phases (e.g. topological superconductors).

## Methods

### Angle Resolved Photoemission Spectroscopy

ARPES measurements on single crystals *Ln*PtBi (*Ln*=Lu, Y) were performed at beamline 10.0.1 of the Advanced Light Source (ALS) at Lawrence Berkeley National Laboratory, USA and beamline I05 of the Diamond Light Source (DLS), UK. The measurement pressure was kept below $3\times10^{-11}$/$8\times10^{-11}$ Torr in ALS/DLS, and data were recorded by Scienta R4000 analyzers at 20K sample temperature at both facilities. The total convolved energy and angle resolutions were 16meV/30meV and 0.2°/0.2° at ALS/DLS, respectively.

**Ab-initio calculations**

To simulate a surface, a 54-atomic-layer thick slab model was used with a vacuum more than 10 Å to diminish the coupling between the top and bottom surfaces. The *ab-initio* calculations were performed within the framework of the density-functional theory (DFT) and generalized gradient approximation[39, 40]. In the bulk calculations, the DFT Bloch wave functions were projected to Wannier functions[36], Ln-*d*, Pt-*sd*, and Bi-*p* atomic like orbitals. In a half-infinite surface model, we projected the Green function of the bulk to the surface unit cell and obtained the surface LDOS based on the Wannier functions.

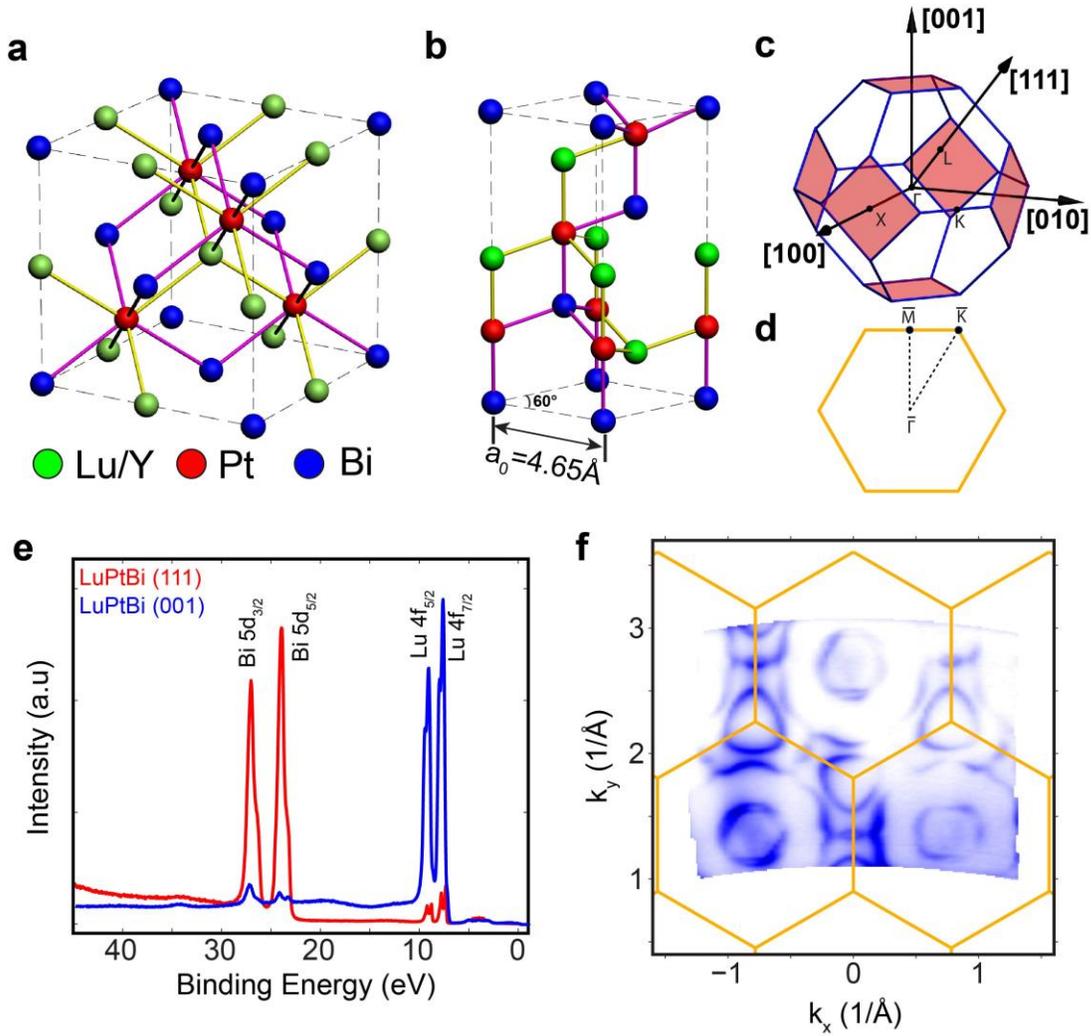

**Figure 1 Crystal structure of *Ln*PtBi and cleavage surface measured by ARPES.** (a) Crystal structure of half-Heusler alloy *Ln*PtBi crystal shows a composite of zinc-blend and rocksalt lattices. (b) Unit-cell of *Ln*PtBi at the (111) cleavage surface shows the stacking of triangular *Ln*, Pt and Bi layers. $a_0$ is the in-plane lattice constant of the (111) surface unit-cell. (c) Bulk BZ of *Ln*PtBi with high symmetry points labelled. Arrows and shaded surfaces indicate the projection to [100], [010], [001] directions. (d) Surface BZ in the [111] direction with the high symmetry points labelled. (e) Core level photoemission spectrum on LuPtBi (111) and (001) surfaces clearly shows the characteristic Lu *4f* and Bi *5d* doublets. These spectra are measured with 75eV and 215eV photons, repectively. (f) Broad FS map of LuPtBi covering 5 BZs, confirming the shape and size of the surface BZ (overlaid yellow hexagons) on the (111) cleave plane. The uneven intensity of the FS at different BZs results from the matrix element effect.

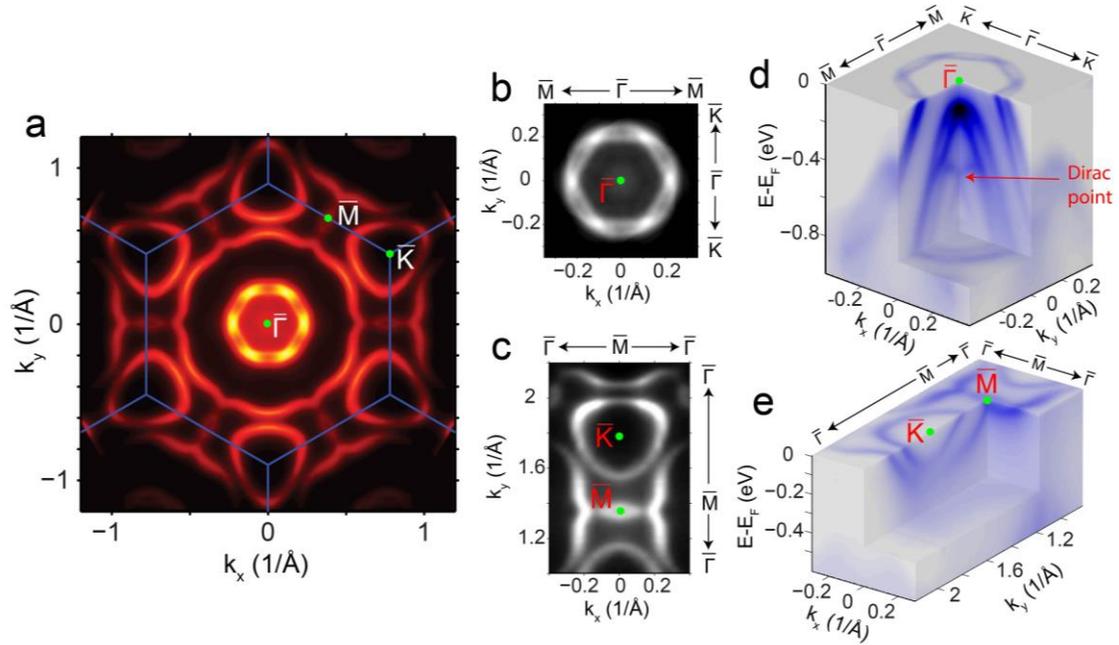

**Figure 2 General electronic structure of LuPtBi (111) surface.** (a) Fermi surface maps of Bi-terminated LuPtBi (111) surface. Blue lines denote the surface BZ with high symmetry points labelled. The data has been symmetrized according to the crystal symmetry. (b)-(c) Zoom-in plot of FS map around the $\bar{\Gamma}$ point (b) and around the $\bar{M}$ and $\bar{K}$ points (c). (d)-(e) Plot of three dimensional electronic structure around the $\bar{\Gamma}$ point (d) and around the $\bar{M}$ and $\bar{K}$ points (e). All data were taken with 65 eV photons.

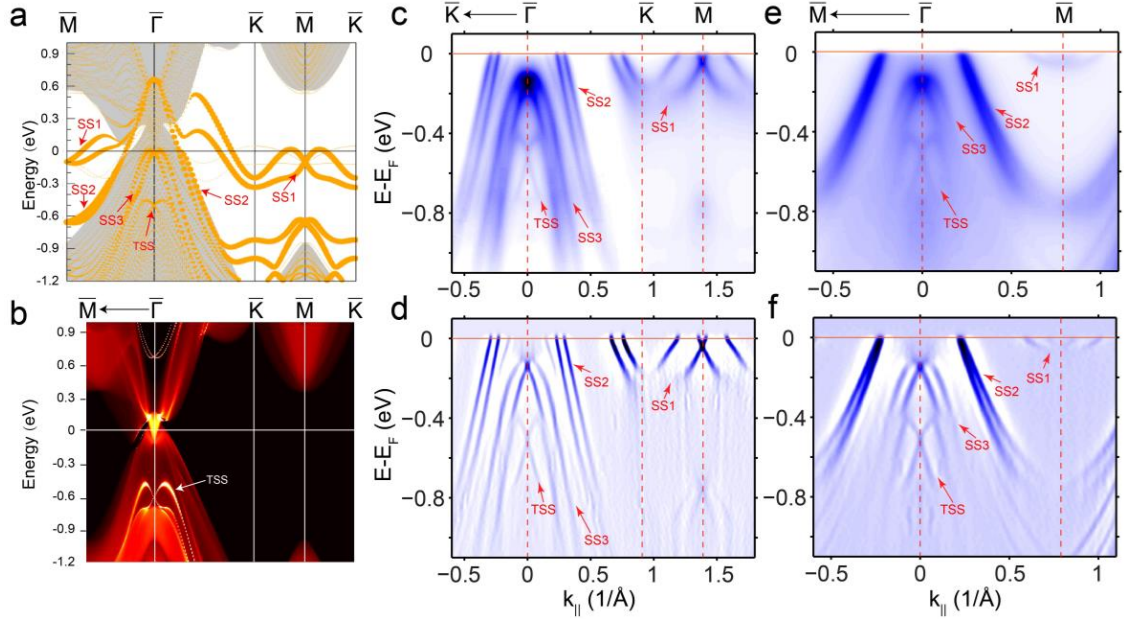

**Figure 3 Observation of the metallic surface state and topological surface state on LuPtBi (111) surface.** (a-b) Calculated band structures of Bi-terminated LuPtBi (111) surface. (a) Result from a slab model calculation, in which the size of filled circles represent the projection to the Bi-terminated surface. Both topologically nontrivial surface state and metallic surface states are captured. (b) Results from a semi-infinite surface that is terminated by Bi. Only topologically nontrivial surface state is revealed by the calculation. (c-d) Photoemission intensity plot (c) and its second-derivative $\frac{\partial^2 I}{\partial E^2}$ plot (d) along the high symmetry $\bar{\Gamma}$-$\bar{K}$-$\bar{M}$ directions. (e-f) Photoemission intensity plot (e) and its second-derivative $\frac{\partial^2 I}{\partial E^2}$ plot (f) along the high symmetry $\bar{\Gamma}$-$\bar{M}$ directions. SS: topologically trivial metallic surface state due to the dangling bonds on sample surface. TSS: topologically non-trivial surface state. All data taken with 65 eV photons.

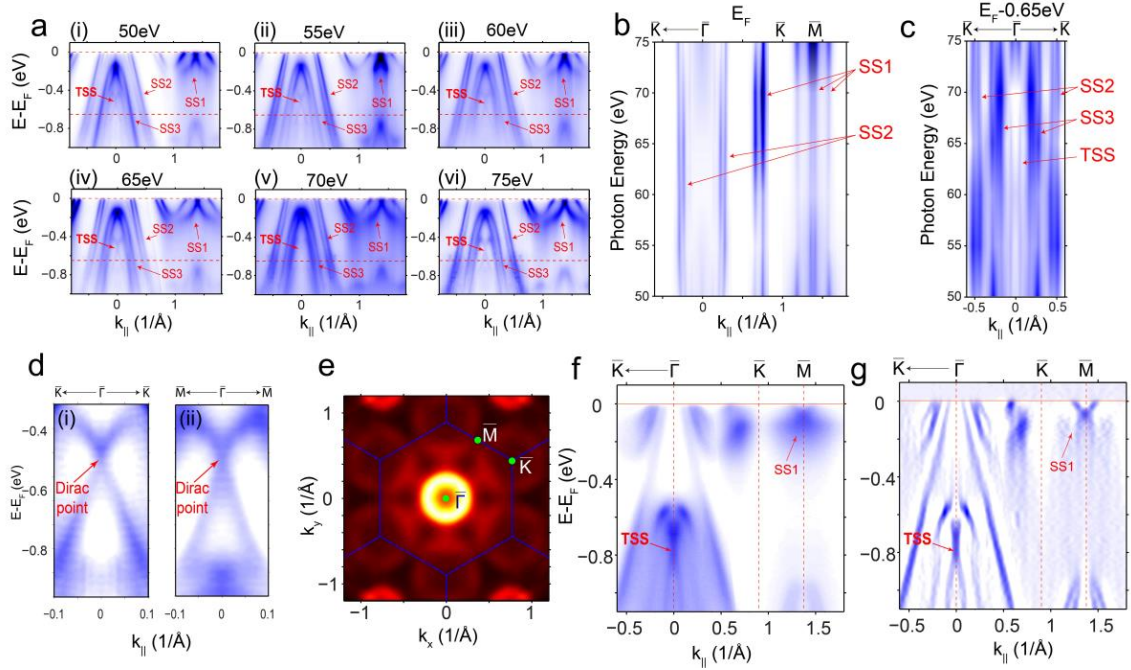

**Figure 4 Photon energy dependence and Dirac-fermion behaviour of the topological surface state.** (a) Photoemission intensity plots along the high symmetry $\bar{\Gamma}$-$\bar{K}$-$\bar{M}$ direction with photon energies from 50eV to 75eV. Two red dotted lines marked the energy where individual MDC is taken to generate the plots in (b) and (c). (b)-(c) Intensity plot of the MDCs taken at $E_F$ (b) and $E_F$-0.65eV (c) at different photon energies. The MDC peak positions of all the observed bands are labelled (SS1~SS3 and TSS). (d) Zoom-in intensity plots of the TSS along the $\bar{\Gamma}$-$\bar{K}$ direction (panel i) and the $\bar{\Gamma}$-$\bar{M}$ direction (panel ii). Dirac points of the TSS are labeled. Data are taken with 60 eV photons. (e-g) Measured electronic structure of (111) surface of YPtBi. (e) The Fermi surface mapping of (111) surface of YPtBi with the hexagonal BZ (overlaid blue lines). The data has been symmetrized according to the crystal symmetry. (f-g) Photoemission intensity (f) and its second-derivative $\frac{\partial^2 I}{\partial E^2}$ plot (g) along the high symmetry $\bar{\Gamma}$-$\bar{K}$-$\bar{M}$ direction. Data in (e-g) are taken with 70eV photons at T=20K. SS: topologically trivial metallic surface state due to the dangling bonds on sample surface. TSS: topologically non-trivial surface state.

**References:**

1. Qi, X., Zhang, S. -C., Topological insulators and superconductors. *Reviews of Modern Physics* **83**, 1057-1110 (2011).

2. Hasan, M. Z., Kane, C. L., Colloquium: Topological insulators. *Reviews of Modern Physics* **82**, 3045(2010).

3. Fisk, Z., *et al.*, Massive electron state in YbBiPt. *Physical Review Letters* **67**, 3310 (1991).

4. Canfield, P. C. *et al.*, Magnetism and heavy fermion - like behavior in the RBiPt series. *J. Appl. Phys.* **70**, 5800 (1991).

5. Butch, N. P., Syers, P., Kirshenbaum, K., Hope, A. P., Paglione, J., Superconductivity in the topological semimetal YPtBi. *Physical Review B* **84**, 220504 (2011).

6. Goll, G., *et al.*, Thermodynamic and transport properties of the non-centrosymmetric superconductor LaBiPt. *Physica B: Condensed Matter* **403**, 1065-1067 (2008).

7. Shekhar, C., Ouardi, S., Nayak, A. K., Fecher, G. H., Schnelle, W., Felser, C., Ultrahigh mobility and nonsaturating magnetoresistance in Heusler topological insulators. *Physical Review B* **86**, 155314 (2012).

8. Tafti, F.F., *et al.*, Superconductivity in the noncentrosymmetric half-Heusler compound LuPtBi: A candidate for topological superconductivity. *Physical Review B* **87**, 184504 (2013).

9. Hou, Z., *et al.*, Extremely High Electron Mobility and Large Magnetoresistance in the Half-Heusler Semimetal LuPtBi. *Physical Review B* **92**, 235134 (2015).

10. Chadov, S., Qi, X., Kübler, J., Fecher, G. H., Felser, C., Zhang, S.-C., Tunable multifunctional topological insulators in ternary Heusler compounds. *Nat. Mater.* **9**, 541-545 (2010).

11. Lin, H., *et al.*, Half-Heusler ternary compounds as new multifunctional experimental platforms for topological quantum phenomena. *Nat. Mater.* **9**, 546-549 (2010).

12. Xiao, D., *et al.*, Half-Heusler Compounds as a New Class of Three-Dimensional Topological Insulators. *Physical Review Letters* **105**, 096404 (2010).

13. Ruan, J., Jian, S. -K., Yao, H., Zhang, H. J., Zhang, S. –C., Xing, D., Symmetry-protected ideal Weyl semimetal in HgTe-class materials. arXiv:1511.08284 (2015).

14. Ando, Y. Topological Insulator Materials. *Journal of the Physical Society of Japan*, **82**, 102001 (2013).

15. Yan, B., Zhang, S. –C., Topological materials. *Rep. on Prog. in Phys*, **75**, 096501 (2012).

16. Hsieh, T. H., Lin, H., Liu, J., Duan, W., Bansil, A., Fu, L., Topological crystalline insulators in the SnTe material class. *Nat. Commun.* **3**, 982 (2012).

17. Dziawa, P., *et al.*, Topological crystalline insulator states in $Pb_{1-x}Sn_xSe$. *Nat. Mater.* **11**, 1023-1027 (2012).

18. Liu, Z. K., *et al.*, Discovery of a Three-Dimensional Topological Dirac Semimetal, $Na_3Bi$. *Science* **343**, 864-867 (2014).

19. Liu, Z. K., *et al.*, A stable three-dimensional topological Dirac semimetal $Cd_3As_2$. *Nat. Mater.* **13**, 677-681 (2014).

20. Xu, S. –Y., *et al.*, Discovery of a Weyl fermion semimetal and topological Fermi arcs. *Science* **349**, 613-617 (2015).

21. Lv, B. Q., *et al.,* Experimental Discovery of Weyl Semimetal TaAs. *Physical Review X* **5**, 031013 (2015).

22. Yang, L. X., *et al.*, Weyl semimetal phase in the non-centrosymmetric compound TaAs. *Nat. Phys.* **11**, 728-732 (2015).

23. Aoki, Y., Sato, H. R., Sugawara, H., Sato, H., Anomalous magnetic properties of Heusler superconductor $YbPd_2Sn$. *Physica C: Superconductivity* **333**, 187-194 (2000).

24. Xu, G., *et al.*, Weak Antilocalization Effect and Noncentrosymmetric Superconductivity in a Topologically Nontrivial Semimetal LuPdBi. *Scientific Reports* **4**, 5709 (2014).

25. Pan, Y., *et al.*, Superconductivity and magnetic order in the noncentrosymmetric half-Heusler compound ErPdBi. *Europhysics Letters* **104**, 27001 (2013).

26. Yan, B., de Visser, A., Half-Heusler Topological Insulator, *MRS Bulletin* **39**, 859-866 (2014).

27. Feng, W., Wen, J., Zhou, J., Xiao, D., Yao, Y., First-principles calculation of topological invariants within the FP-LAPW formalism. *Computer Physics Communications* **183**, 1849-1859 (2012).

28. Fu, L., Berg, E., Odd-Parity Topological Superconductors: Theory and Application to $Cu_xBi_2Se_3$. *Physical Review Letters* **105**, 097001(2010).

29. Liu, C., *et al.*, Metallic surface electronic state in half-Heusler compounds $R$PtBi ($R$ = Lu, Dy, Gd), *Physical Review B* **83**, 205133 (2011).

30. Chandra, S., *et al.*, Large linear magnetoresistance and weak anti-localization in Y(Lu)PtBi topological insulators. arXiv:1502.00604 (2015).

31. Chen, Y. L., Studies on the electronic structures of three-dimensional topological insulators by angle resolved photoemission spectroscopy. *Frontiers of Physics* **7**, 175-192 (2012).

32. Chu, R. –L., Shan, W. –Y., Lu, J., Shen, S. -Q., Surface and edge states in topological semimetals. *Physical Review B* **83**, 075110 (2011).

33. Wu, S. –C., Yan, B. H., Felser, C., *Ab initio* study of topological surface states of strained HgTe. *Europhysics Letters* **107**, 57006 (2014).

34. Brüne, C., *et al.*, Quantum Hall Effect from the Topological Surface States of Strained Bulk HgTe. *Physical Review Letters* **106**, 126803 (2011).

35. Sancho, M. P. L., Sancho, J. M. L., Rubio, J., Quick iterative scheme for the calculation of transfer matrices: application to Mo (100). *Journal of Physics F: Metal Physics* **14**, 1205 (1984).

36. Mostofi, A. A., Yates, J. R., Lee, Y. –S., Souza, I., Vanderbilt, D., Marzari, N., wannier90: A tool for obtaining maximally-localised Wannier functions. *Computer Physics Communications*, **178**, 685-699 (2008).

37. Qi, X. –L., Li, R., Zang, J., Zhang, S. -C., Inducing a Magnetic Monopole with Topological Surface States. *Science* **323**, 1184-1187 (2009).

38. Li, R., Wang, J., Qi, X. –L., Zhang, S. -C. Dynamical axion field in topological magnetic insulators. *Nature Physics* **6**, 284-288 (2010).

39. Kresse, G., Hafner, J., *Ab initio* molecular dynamics for liquid metals. *Physical Review B* **47**, 558-561 (1993).

40. Perdew, J. P., Burke, K., Ernzerhof, M., Generalized Gradient Approximation Made Simple. *Physical Review Letters* **77**, 3865-3868 (1996).